\documentclass[journal]{IEEEtran}

\usepackage{graphicx}
\usepackage{subfigure}
\usepackage{epstopdf}
\usepackage{multirow}
\usepackage{diagbox}
\usepackage{textcomp,booktabs}
\usepackage[usenames,dvipsnames]{color}
\usepackage{colortbl}
\usepackage{slashbox}
\usepackage{makecell}
\newcommand{\tabincell}[2]{\begin{tabular}{@{}#1@{}}#2\end{tabular}}
\definecolor{mygray}{gray}{.9}
\definecolor{mypink}{rgb}{.99,.91,.95}
\definecolor{mycyan}{cmyk}{.3,0,0,0}

\ifCLASSINFOpdf

\else

\fi

\begin{document}

\title{ECD: An Edge Content Delivery and Update Framework in Mobile Edge Computing}

\author{Shangguang~Wang,~\IEEEmembership{Senior~Member,~IEEE,}
        Chuntao~Ding,
        Ning Zhang,~\IEEEmembership{Member,~IEEE,}\\
        Nan Cheng,~\IEEEmembership{Member,~IEEE,}
        Jie~Huang,
        Ying~Liu
        %and~Jane~Doe,~\IEEEmembership{Life~Fellow,~IEEE}% <-this % stops a space

\thanks{Shangguang Wang, Chuntao Ding, Jie Huang and Ying Liu are with Beijing University of Posts and Telecommunications, Beijing, China.}
\thanks{
Ning Zhang is with Texas A\&M University at Corpus Christi, Texas, USA.}
\thanks{
Nan Cheng is with University of Waterloo, Ontario, Canada.}% <-this % stops a space

}

\maketitle

\begin{abstract}
This article proposes an edge content delivery framework (ECD) based on mobile edge computing in the era of Internet of Things (IOT), to alleviate the load of the core network and improve the quality of experience (QoE) of mobile users. Considering mobile devices become both the content consumers and providers, and majority of the contents are unnecessary to be uploaded to the cloud datacenter, at the network edge, we deploy a content server to store the raw contents generated from the mobile users, and a cache pool to store the contents that are frequently requested by mobile users in the ECD. The cache pools are ranked and high ranked cache pools will store contents with higher popularity. Furthermore, we propose edge content delivery scheme and edge content update scheme, based on content popularity and cache pool ranking. The content delivery scheme is to efficiently deliver contents to mobile users, while the edge content update scheme is to mitigate the content generated by users to appropriate cache pools based on its request frequently and cache poor ranking. The edge content delivery is completely different from the content delivery network and can further reduce the load on the core network. In addition, because the top ranking cache pools are prioritized for higher priority contents and the cache pools are prioritized for higher priority contents and the cache pools are in proximity to the mobile users, the immediately interactive response between mobile users and cache pools can be achieved. A representative case study of ECD is provided and open research issues are discussed.
\end{abstract}


\IEEEpeerreviewmaketitle

\section{Introduction}
\IEEEPARstart{W}{ITH} the emerging Internet of Things (IoT), we have witnessed the proliferation of mobile devices, such as smart phones, laptops, connected vehicles, and sensors. Those end devices are usually resource-constrained. In contrast, cloud datacenter has unlimited resources~\cite{Fernando@Mobile}. Offloading the computation and contents to the cloud datacenter can address the tension between resource-hungry applications and resource-constrained mobile devices. However, with the proliferation of mobile devices, massive devices are connected and generate tremendous data traffic. According to Cisco Visual Networking Index~\cite{CiscoWhitePaper}, mobile devices and connections will grow up to 11.6 billion by 2021, and the share of smart devices and connections will also increase from 46\% in 2016 to 75\% in 2021. As a result, uploading all the contents to the remote cloud datacenter will consume massive amounts of network bandwidth and increase the load on the core network. In addition, because of the distance from mobile users to cloud datacenter, long latency is inevitable and degrade service performance.

To alleviate the disadvantages, researchers from both academic and industry are looking to push contents and infrastructure close to mobile users to alleviate the core network traffic and improve the quality of experience (QoE) of users~\cite{Florin@Understanding,Sun@Beyond}. Content Delivery Network(CDN) is proposed to optimize the network traffic and improve the QoE of users~\cite{JCaching CDN,CDNI}. In CDN, content providers upload all the contents generated by users to the cloud datacenter directly. If a content is requested by mobile devices, CDN first delivers the content from cloud datacenter to proxy servers. Then, the mobile devices can obtain the required content from the proxy servers rather than cloud datacenter, which can reduce the load on the core network and improve the user's QoE to some extent. However, the distance between proxy servers and mobile devices is still far away, which cannot satisfy the requirements of some real-time applications. In addition, mobile devices have the dual role of both content generators\/providers and content consumers. If all the generated contents are uploaded to the cloud datacenter, it will consume massive bandwidth and swamp the core network. As a matter of fact, the majority of contents are unnecessary to be uploaded to the cloud datacenter, because only minority of the contents can be requested frequently in reality.

To complement CDN, a novel computing paradigm, called mobile edge computing (MEC) has been proposed~\cite{MECSurvery,MECC,CollaborativeMECSurvery,Taleb@Multiaccess}. By deploying edge servers in the pervasive radio access networks, MEC can provide cloud capabilities (e.g., computing, storage, and caching) in close proximity to mobile users, without the need to offload computation and contents to the cloud datacenter. Since the edge servers are near to the mobile devices, the immediately interactive response can be met.

Fueled by the potential capabilities of MEC, we propose an edge content delivery framework (ECD). The proposed ECD is completely different from CDN. It consists of three layers: end layer, edge layer and cloud layer. In the end layer, diverse mobile devices can generate diverse contents and consume\/request different contents. In the edge layer, we deploy content servers and cache pools in the access network, which store the raw contents uploaded from the mobile users and the contents that are frequently requested by mobile users, respectively. As a result, mobile devices only need to upload all the generated contents to the content server rather than the remote cloud datacenter. Thus, the load on the core network will be significantly reduced. We propose an edge content delivery scheme, where we prioritize the top ranking cache pools to store higher priority contents. Furthermore, we also propose an edge content update scheme to upload the popular contents to different ranked cache pools and the cloud datacenter. The raw content stored in content servers can only be mitigated to different ranked cache pools and the cloud datacenter, if its request frequently is keeping increasing.

The remainder of this article is organized as follows. In the following section, we propose an edge content delivery framework. Then, we provide a comparison between ECD and CDN in various features. Following that, we describe a typical case study to illustrate the benefits of ECD. Then, we highlight some new challenges that need to be tackled. Finally, we draw our conclusion in the final section.

\section{ECD Framework}
The proposed ECD framework consists of three layers, end layer, edge layer and cloud layer. In ECD, the edge content delivery and update storages are performed in a collaborative manner. In the following, we discuss the ECD in detail from three aspects, component, edge content delivery and edge content update, respectively.

\subsection{Component}
The main characteristics and functions of each layer are as follows.

\subsubsection{End layer} The end layer consists of various mobile devices, such as Google Glass, Smartphone, Laptop, Vehicle, and so on. These devices are not only generating different contents but also consuming contents. Nevertheless, mobile devices are in general resource-constrained, having limited computation capability and storage capacity. It is necessary to upload the collected contents to other servers with powerful computing capability and enough storage capacity.

\subsubsection{Edge layer}
The edge layer is between end layer and cloud layer, and consists of two parts, base station(BS) and edge server. The role of the BS is to communicate with the mobile devices, the cloud datacenter and other BSs. The edge server also consists of two parts, content server and cache pool\footnote{in this article, base station, content server and cache pool are bound together}. Content server has unlimited storage capacity and is used to store all the raw contents uploaded from mobile users while cache pool has also limited storage capacity and is used to store the contents that are frequently requested by users. For example, if a mobile user wants to upload a video, he first uploads the video to the content server through LTE base station. If the video is frequently requested, the video will be migrated to the cache pool based on a cache content update scheme. Otherwise, it will be stored in the content server.
\begin{figure*}
   \centering
   \begin{center}
     \includegraphics[width=1\linewidth]{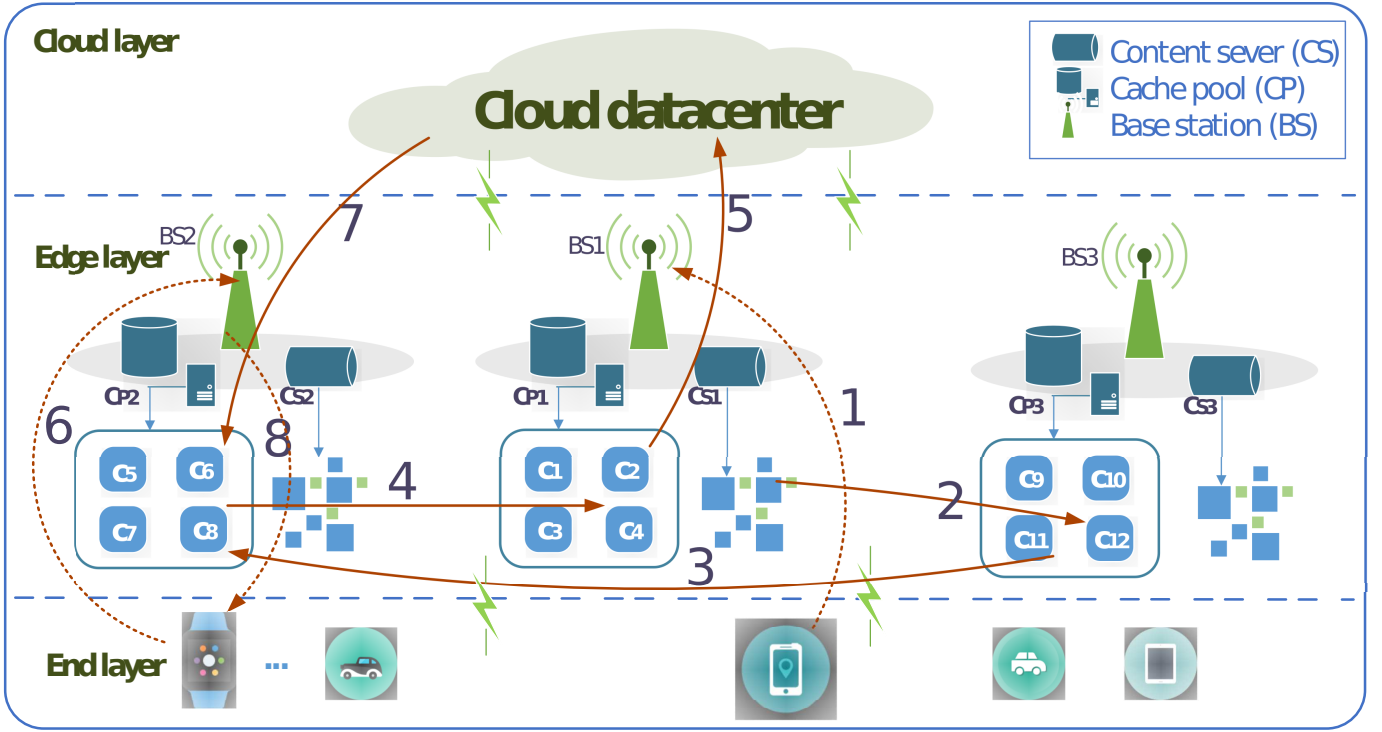}
   \caption{Edge content update process. The detailed process is as follow: 1. a mobile user uploads a content such as $c_{14}$ to $C_{S1}$ through $BS1$; 2. $C_{S1}$ copies a copy of $c_{14}$ and migrates it to $C_{P3}$; 3. $C_{P3}$ migrates $c_{14}$ to $C_{P2}$; 4. $C_{P2}$ migrates $c_{14}$ to $C_{P1}$; 5. $C_{P1}$ uploads $c_{14}$ to the cloud datacenter; 6. other users request $c_{14}$ from $C_{P2}$ through $BS2$; 7. the cloud datacenter delivers $c_{14}$ to $C_{P2}$; 8. $C_{P2}$ provides $c_{14}$ service to mobile users.}
   \label{fig:2}
   \end{center}
\end{figure*}

\subsubsection{Cloud layer}
In the cloud layer, the cloud datacenter is considered to have unlimited computation capability and unlimited storage capacity. However, uploading all the collected contents from mobile devices to a cloud datacenter consumes massive bandwidth and has long network latency. In order to reduce the load on the core network as well as latency, it is necessary to avoid excessive content uploading. As a matter of fact, according to the Pareto principle, the majority of contents are unnecessary to be uploaded to cloud datacenter in reality. Therefore, in the proposed framework, the cloud datacenter only stores and delivers popular contents.

\subsection{Edge content delivery}
Edge content delivery scheme is to efficiently deliver contents to mobile users by storing popular contents in appropriate cache pools. It works as follows:

\textbf{\emph{Initialization:}} We assume that all the contents are stored in the cloud datacenter, the cache pools have not yet deployed any contents, and the location of all the base stations are fixed. The base stations are considered as vertices, the cost of communications between base stations is considered as weights. All the base stations form a weighted complete graph.  For the sake of simplicity, the contents are forbidden to be duplicated in the initial stage.

\textbf{\emph{Where to deliver:}} The Floyd-Warshall algorithm is employed to search optimal solution since it can search the lengths of the shortest paths between all pairs of vertices. By operating the Floyd-Warshall algorithm, we obtain a list sorted by cache pool ranking. The higher the ranking is, the better the location of the cache pool is in the weighted complete graph.

\textbf{\emph{Which to deliver:}} Since all the contents are stored in the cloud datacenter in the initial stage, the frequency that the content is requested can be known by the cloud. For simplicity, we assign the priority of the content based on the frequency it is requests (e.g., in a certain time range, such as within six months). The more the number of requested, the higher the priority. Here, we prioritize the delivery of higher priority contents to the cache pools.

\textbf{\emph{Delivery strategy:}} For better edge content delivery, we choose the top-$K$ cache pools as the main cache pools, which store popular contents. For the top-$K$ cache pools, we first deliver the content with the highest priority to the optimal cache pool, and then the content with the second highest priority, until the optimal cache pool is full. Then, we deliver the contents to the suboptimal cache pool, and so on, until all the top-$K$ cache pools are full. The remaining cache pools also follow the rules. Please note, in order to facilitate the content update, we only use about $\varepsilon$ times of the storage capacity of the remaining cache pools to store contents.
\subsection{Edge content update}
The edge content update scheme is to update the contents in the cache pools and datacenters. For instance, the raw content stored in content servers can be mitigated to cache pools if it becomes popular (e.g., by requested frequently). Moreover, the content stored in cache pools can only be mitigated to different ranked cache pools and the cloud datacenter, according to the time-varying content popularity.

To clearly describe the update process, we give an example to illustrate it, as shown in Figure 1. In Figure 1, $C_{P1}$, $C_{P2}$, $C_{P3}$ represent the cache pools and the ranking of them is $C_{P1}$$>$$C_{P2}$$>$$C_{P3}$, and $C_{P1}$ is the top-$K$ cache pool (here, $K$=1). $BS1, BS2, BS3$ represent the base stations, $C_{S1}, C_{S2}, C_{S3}$ are the content servers, $c_{1}, c_{2},\cdots, c_{12}$ represent the contents, and
$r_{c1}$, $r_{c2}$, $\cdots$, $r_{c12}$ represent the number of times the contents have been requested, and $r_{c1}> r_{c2}>,\cdots,>r_{c12}$. The contents on the top-$K$ cache pools are popular contents, here, $c_{1}, c_{2},c_{3}, c_{4}$ are popular contents. We consider the following three situations:

\subsubsection{User's request is met on the cache pools}
If the contents requested by the mobile users are cached in the cache pools, the cache pools will transmit the content services to the mobile users. Suppose that the request frequency of a content such as $c_{10}$ keep increasing, the edge content update scheme will be performed as follows:

\textbf{Case 1:} With the number of times $c_{10}$ has been requested increased until $r_{c10}$ more than $r_{c8}$ $\delta$ times, where $\delta$ is a constant, $C_{P3}$ checks whether $C_{P2}$ has enough storage capacity for $c_{10}$. If yes, $C_{P3}$ migrates $c_{10}$ to $C_{P2}$. Otherwise, $C_{P2}$ migrates the least requested contents, i.e. $c_{8}$ or even $c_{7}$ to $C_{P3}$ for the storage of $c_{10}$. Simultaneously, $C_{P2}$ checks whether $C_{P3}$ has enough storage capacity for $c_{8}$. If yes, $C_{P2}$ migrates $c_{8}$ to $C_{P3}$. If not, $C_{P3}$ deletes the least requested contents $c_{12}$ or even $c_{11}$.

\textbf{Case 2:} For the contents stored in top-$K$ server $C_{P1}$, $C_{P1}$ counts the number of requested for the contents through each cache pool. If the number of requested of $c_{2}$ through ${BS2}$ more than $\varepsilon$ times of the total number of requested and $C_{P2}$ has enough storage capacity for $c_{2}$, cloud datacenter delivers $c_{2}$ to $C_{P2}$ directly. Otherwise, $C_{P2}$ has not enough storage capacity currently, similar to case 1, $C_{P2}$ migrates $c_{8}$ or even $c_{7}$ to $C_{P3}$ then.

\subsubsection{User's request is not met on the cache pools}
When mobile users request a new content, e.g., $c_{13}$, which is not stored in the cache pools, the cloud datacenter will check whether the cache pool with the lowest ranking (i.e. $C_{P3}$) has enough storage capacity for $c_{13}$. If yes, the cloud datacenter delivers $c_{13}$ to $C_{P3}$. Otherwise, $C_{P3}$ deletes the least requested contents $c_{12}$ or even $c_{11}$, until $C_{P3}$ has enough storage capacity. After that, the cloud datacenter delivers $c_{13}$ to $C_{P3}$.
\begin{table*}[!htp]
\renewcommand{\arraystretch}{2.0}
\caption{Comparison of features: ECD vs. CDN}
\centering  % 表居中
\begin{tabular}{>{\columncolor{mycyan}\sf }l|>{\columncolor{mygray}\sf }c|>{\columncolor{mygray}\sf }c}

\rowcolor{mycyan}
 &\bf{ECD} &\bf{CDN}\\
\Xhline{2\arrayrulewidth}
\bf{Distance to mobile users} &\tabincell{c}{Small (tens to hundreds of meters)} &\tabincell{c}{Large (tens of kilometers)} \\ \Xhline{1\arrayrulewidth}
\bf{Upload Strategy} &\tabincell{c}{Upload all the captured contents to the content server and \\only upload popular contents to the cloud datacenter} &\tabincell{c}{Upload all the captured contents to the cloud \\datacenter directly}   \\ \hline
\bf{The load on the core network} &\tabincell{c}{Light load} &\tabincell{c}{Heavy load}   \\ \hline
\bf{Delivery Strategy} &\tabincell{c}{Only deliver popular contents to\\ the cache pools} &\tabincell{c}{Deliver all the requested contents to \\the proxy servers}   \\ \hline
\bf{Peer communication} &\tabincell{c}{Autonomic communication with data transmission} &\tabincell{c}{Without autonomic communication}   \\ \hline

\bf{Backhaul usage} &\tabincell{c}{Infrequent use (alleviate congestion)} &\tabincell{c}{Frequent use (can cause congestion)} \\ \hline
%\bf{Location awareness}&Yes &N/A \\ \hline
\bf{Collaborative Decision Making} & Yes &N/A \\
\end{tabular}
\end{table*}
\subsubsection{User requests to upload a content}
When a mobile user wants to upload a content, such as $c_{14}$, she uploads $c_{14}$ to her content server $C_{S1}$, and the cache pool $C_{P1}$ adds the description of $c14$. Thus, mobile users can request the content by the description of the content stored in $C_{P1}$. If the request frequency $r_{c14}$ is greater than $r_{c12}$ by $\delta$ times and $C_{P3}$ has enough storage capacity, $C_{S1}$ copies a copy of $c_{14}$ and migrates it to $C_{P3}$. If there is no more storage capacity for $c_{14}$, $C_{P3}$ deletes $c_{12}$ or even $c_{11}$. If and only if $c_{14}$ is migrated to $C_{P1}$, $C_{P1}$ copies a copy of $c_{14}$ and uploads it to the cloud datacenter.

\section{ECD vs. CDN}
To better understand the differences between ECD and CDN, we summarize the comparison between ECD and CDN in various aspects in Table I. In addition, we explain three typical differences between ECD and CDN in the following~\cite{MECSurvery,Rimal@Mobile}.

\textbf{Distance to mobile users: } The communication delay can be significantly reduced since the distance between edge servers and mobile users is smaller than the distance between proxy servers and mobile users. In ECD, edge servers are deployed in close proximity to mobile users. In general, the distance between edge servers and mobile users is tens to hundreds of meters. However, in CDN, proxy servers are deployed in multiple locations, often over multiple backbones. The distance from proxy servers to mobile users is tens of kilometers. Therefore, ECD can provide services to mobile users with lower communication latency, which can significantly improve QoE.

\textbf{Collaborative Decision Making: } In ECD, besides storage, cache pools have capabilities of computation and can perform collaborative decision making. For the proxy servers of CDN, when, where and which contents to deliver are all decided by the cloud datacenter~\cite{OverlayCDN}. There is no autonomic cooperation between proxy servers. The tasks of the proxy servers are confined to storage and transmission under the order of cloud datacenter. In contrast, the cache pools have kinds of decision-making capabilities. The communication between the cache pools is maintained by themselves rather than by the cloud datacenter. The advantage of the collaborative decision making is that the cache pools can obtain the requested times of all the contents stored in other cache pools that they can cooperate with each other and facilitate to decide when, where and which contents to migrate.

\textbf{Upload Strategy: }In ECD, for each base station, we deploy a content server and a cache pool. Thus, the mobile users only need to upload all the collected contents to the content server rather than the cloud datacenter, that can significantly reduce the load on the core network. Moreover, in reality, roughly minority of the contents need to upload to the cloud datacenter. In ECD, rather than simply upload the minority of contents to the cloud datacenter directly, we design an edge content delivery strategy to further reduce the load on the core network. In contrast, content providers upload the collected contents to the cloud datacenter directly in CDN, based on Pareto principle, the majority of contents may not be requested by mobile devices frequently. It is unnecessary to upload the contents to the cloud datacenter, this process will consume massive bandwidth and incur network congestion.

\section{Case study}
\begin{figure}
   \centering
   \begin{center}
     \includegraphics*[width=1\linewidth]{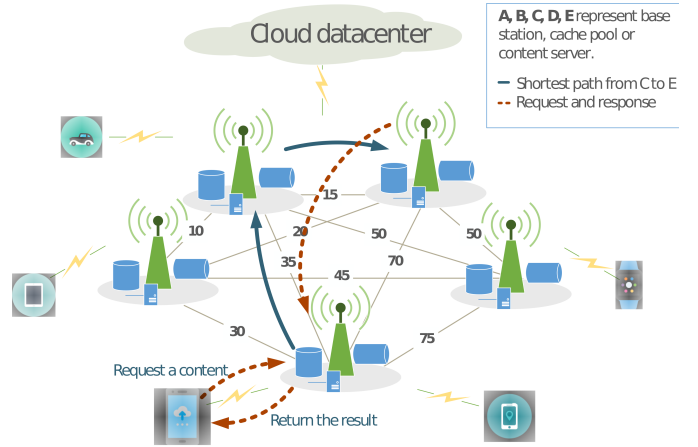}
   \caption{Example of Youtube video request. A mobile user requests a video from E. Based the Floyd-Warshall algorithm, the shortest path from C to E is C-A-E. C transmits the request to E through A and E returns the result to C through A. Finally, C provides the video service to the mobile user.}
   \label{fig:2}
   \end{center}
\end{figure}

To clearly illustrate ECD, Figure 2 gives an example of an analysis of YouTube, since YouTube is the largest video site, and many mobile devices download, watch and share videos from YouTube every day. In Figure 2, A, B, C, D and E represent the base station or content server or cache pool\footnote{since the base station, content server and cache pool are bound together, that A, B, C, D and E can represent them in the context}, and the numbers on the lines are the cost of communication with each other. We assume that there are 1000 videos stored in the cloud datacenter, $v_1,v_2,\cdots,v_{1000}$, respectively.

Based on the edge content delivery scheme, we first determine where to deliver the videos from the cloud datacenter. The Folyd-Warshall algorithm is used to find the ranking of each base station based on the principle of the smaller the cost is, the higher ranking. Table II shows the computational process, the ranking result and the ranking order is B, A, E, C, D.
\begin{table}[!htp]
\renewcommand{\arraystretch}{1.5}
\caption{The ranking of five content servers}
\centering  % 表居中
\begin{tabular}{>{\columncolor{mycyan}\sf }c|>{\columncolor{mygray}\sf }c|>{\columncolor{mygray}\sf }c|>{\columncolor{mygray}\sf }c|>{\columncolor{mygray}\sf }c|>{\columncolor{mygray}\sf }c|>{\columncolor{mygray}\sf }c|>{\columncolor{mygray}\sf }c}
\rowcolor{mycyan}
\bf{Server} &\bf{A} &\bf{B}
& \bf{C}& \bf{D}& \bf{E} & \bf{Total cost} &\bf{Ranking}\\
\Xhline{2\arrayrulewidth}
\bf{A}&0 &10 &35 &50 &15 &110 &2 \\

\bf{B}&10 &0 &30 &45 &20 &105 &1 \\

\bf{C}&35 &30 &0 &75 &50 &190 &4 \\

\bf{D}&65 &45 &75 &0 &50 &235 &5 \\

\bf{E}&15 &20 &50 &50 &0 &135 &3 \\
\end{tabular}
\end{table}

Then, based on the priority of videos, we decide which to deliver, by the number of times the video is requested with a certain time range. The priority of those videos is $v_1>v_2>,\cdots,>v_{1000}$. In the case study, we assume that each cache pool and proxy server can store 10 videos and $\delta$=10\%, $\varepsilon$=1/3.

From Table II, the top-$K$ cache pools (here $K$=2) are B and A. Therefore, the B cache pool stores the top-10 videos, and A cache pool stores the videos with the priority between 10 to 20. And E, C and D only store 3 videos. In summary, in the initial delivery phrase, $v_1$,$\cdots$,$v_{10}$ are delivered to B, $v_{11}$,$\cdots$,$v_{20}$ are delivered to A, $v_{21}, v_{22}, v_{23}$ are delivered to E, $v_{24}, v_{25}, v_{26}$ are delivered to C, $v_{27}, v_{28}, v_{29}$ are delivered to D.
\begin{figure*}[!htp]
\centering     %%% not \center
\subfigure[The number of base stations is 5]{\label{fig:a}\includegraphics[width=0.4\linewidth]{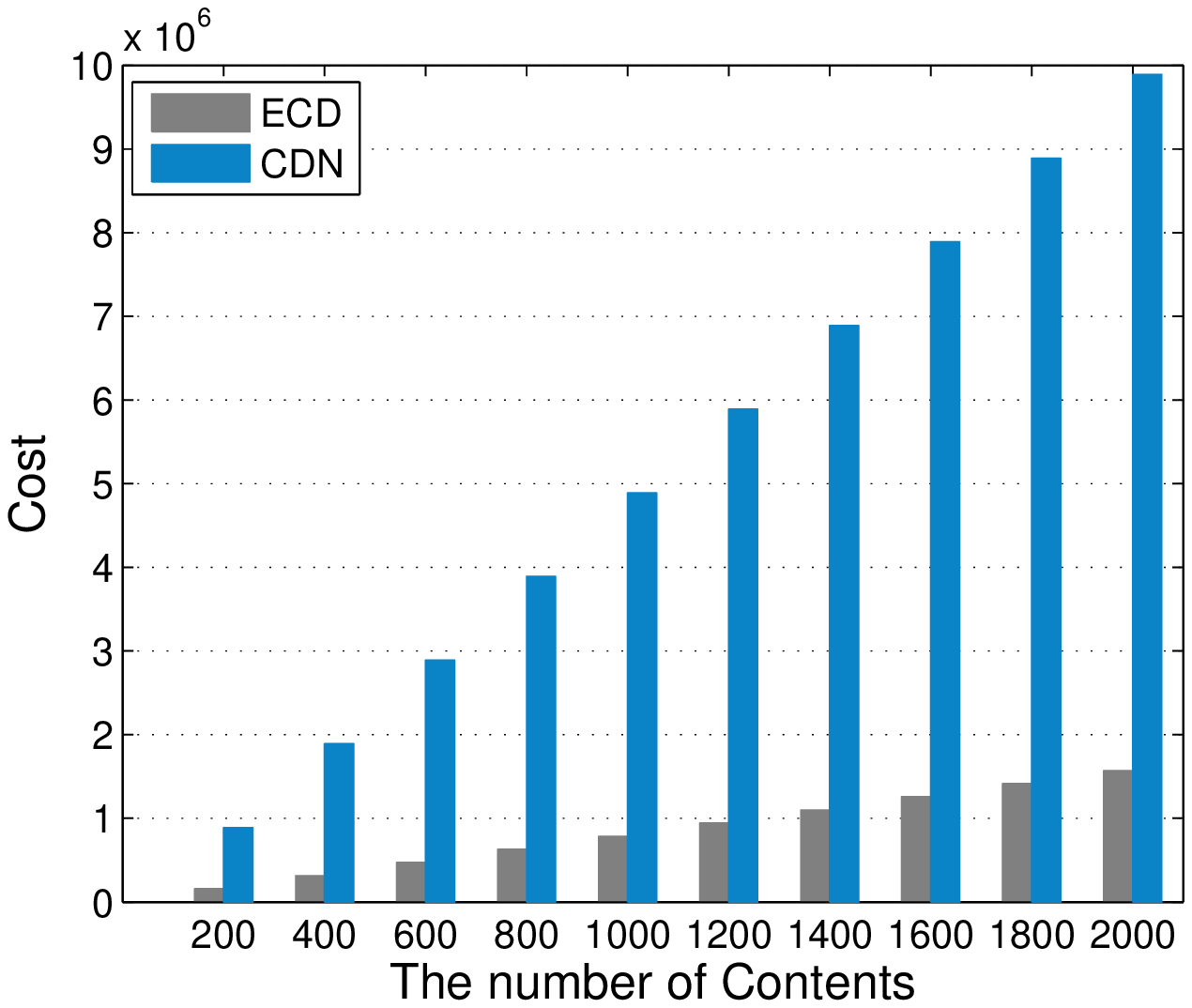}}
\subfigure[The number of base stations is 50]{\label{fig:b}\includegraphics[width=0.4\linewidth]{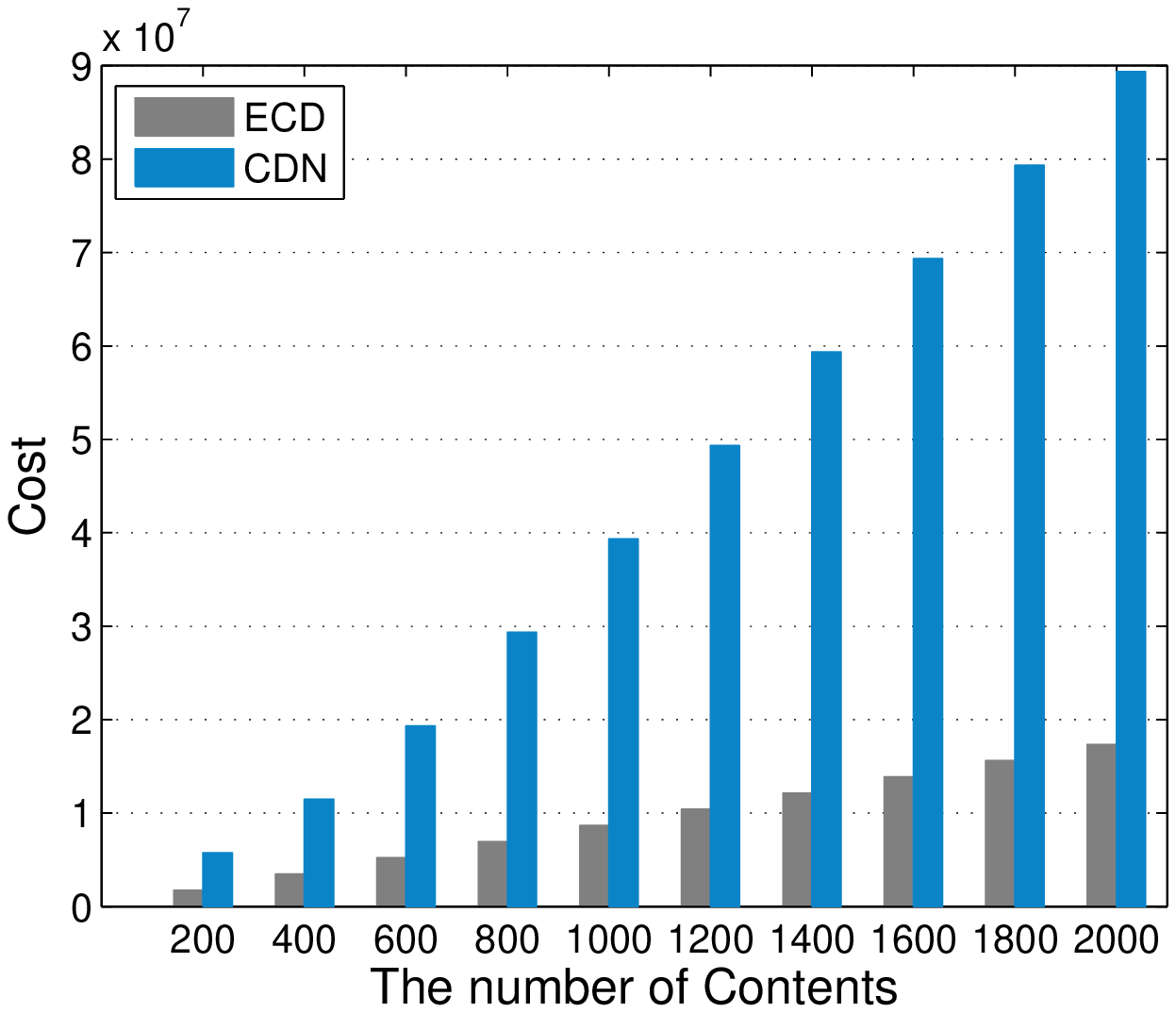}}
\subfigure[The number of base stations is 100]{\label{fig:c}\includegraphics[width=0.4\linewidth]{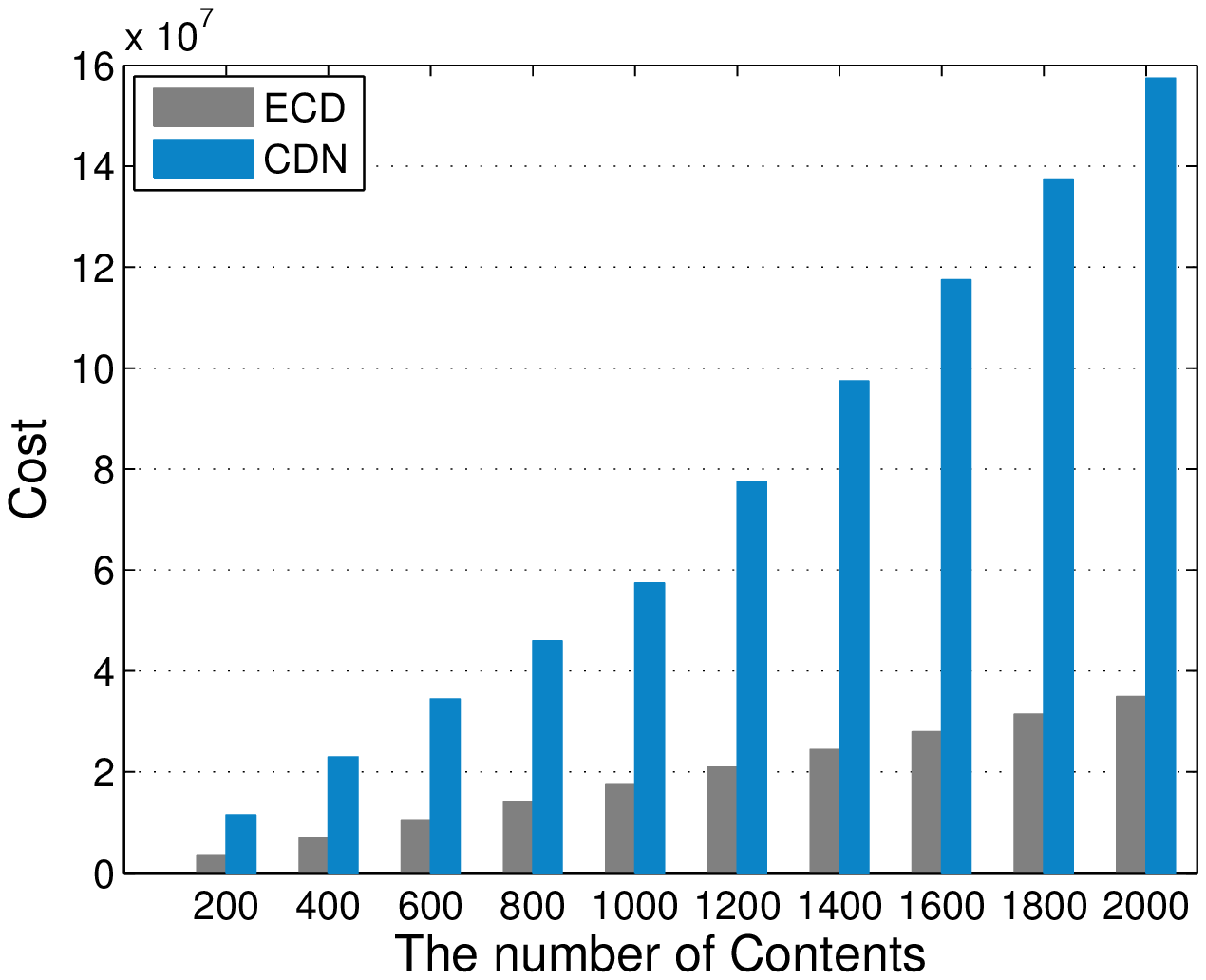}}
\subfigure[The number of base stations is 500]{\label{fig:d}\includegraphics[width=0.4\linewidth]{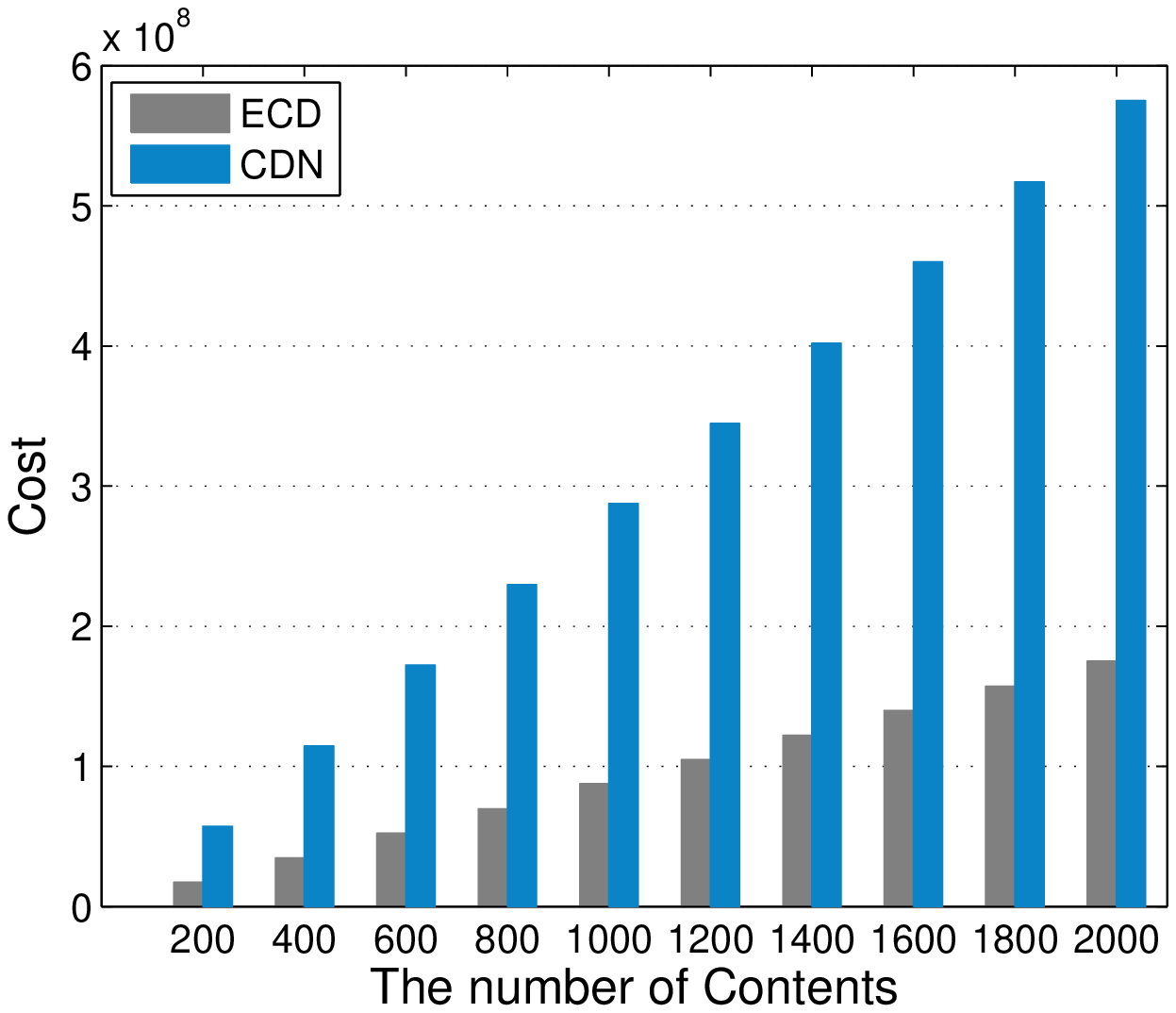}}
\caption{The Cost of ECD and CDN under different numbers of base stations and different numbers of requested contents.}
\end{figure*}

If the requested number of $v_{25}$ in C increased over time, C migrates $v_{25}$ to E when $v_{25}$ is 10\% more than $v_{23}$ in E according to our update strategy. Supporting that E has no more storage capacity for $v_{25}$, E migrates $v_{23}$ to C at the same time. We believe that videos in B and A are popular contents. Thus, as the number of requests to a popular content $v_{11 }$ through C reaches 1/3 of total requests, the cloud datacenter delivers $v_{11}$ to C. Specifically, if a mobile user requests to a new video that is not available on these five servers. Then, we search it in the cloud datacenter and deliver it to D. Additionally, the storage capacity is not enough, D deletes $v_{29}$ directly.

Furthermore, suppose that a mobile user under the coverage of E requests to upload a video. For ECD, the mobile user uploads it to the content server firstly. Other mobile devices are allowed to request the video only with the permission of cache pool E. If the number of requests to the video more than 10\% of $v_{29}$ in D, E copies the video and migrates it to D. D will delete $v_{29}$ if its storage capacity is not enough. If and only if the video is migrated to A or B, which can be uploaded to the cloud datacenter.

Assume the distance between mobile users and the cloud datacenter is 1000. For CDN, the distance between the proxy servers and the mobile users and the distance between the cloud datacenter and the proxy servers are 500. While for ECD, the distance between the edge server and the mobile users is 100, and 900 for the distance between the edge server and the cloud datacenter. Suppose that there are 10 requests for a new video, respectively. CDN delivers these 10 new videos to the proxy servers who request it. Suppose A, B, C, D and E stores two videos each. And ECD delivers these 10 new videos to D, without effect to those popular contents. CDN does not differ from our framework in terms of cost at delivery, both of them are 1000*10. However, after the delivery to the proxy servers, when there are other requests for these 10 videos from mobile users, the cost of CDN is  (110+105+190+235+135)*2*500. And for ECD, the cost is 235*10*100 in the worst case, less than 69.67\% of CDN as well as 105*10*100 for the best case, which is about 86.45\% less than CDN. In summary, the cost of ECD saves about 73.03\% in average, compared with CDN. In addition, when a mobile device requests upload a new video, CDN upload it to the cloud datacenter with a possibility of 100\%. And its cost is 1000*100\% obviously. According to the Pareto Principle, the possibility of uploading to the cloud datacenter is only 20\%. The price is 100*100\%+900*20\% for ECD. In this way, the ECD saves 72\% of the cost.

Furthermore, Fig. 4 shows the cost of ECD and CDN at different numbers of base stations as well as different numbers of requested contents and the cost of ECD is lower than the CDN. The main reason is that in ECD, all the contents stored in the content servers and mobile devices do not need to request contents from the cloud datacenter, which can save a lot of cost. However, in CDN, all the contents are stored in the cloud datacenter, and when the requested contents do not stored in the proxy servers, mobile devices need to request the contents from the cloud datacenter, which will consume a lot of cost, because mobile devices are far away from the cloud datacenter. From Figure 3, it can seen that the cost of ECD always lower than CDN in different situations, which means that ECD can not only reduce the load on the core network but also reduce response time.
 \section{New Challenges}
The proposed edge content delivery framework brings various benefits such as reducing the load on the core network and improving the user's QoE, and at the same time introduces new challenges, highlighted in the following~\cite{Rimal@Mobile}.

\textbf{Network Integration and Coordination: }
Under the diverse potential deployment scenarios over multiple RANs (e.g., WLAN, LTE), integration of MEC network should be valued at both the architectural and protocol levels. Moreover, the cooperation between front-end and back-haul segments of converged networks in 5G is also a vital issue.

\textbf{Resource Management: }
In practice, storage and computing resources of cache pools are limited and can only support a restricted number of services. When the limited amount of shared resources should be allocated to meet dynamic needs of mobile devices, the complexity of allocation strategy increases a lot. So it's really a challenge to design a resource management scheme for the network with high QoE.

\textbf{Cloud datacenter and Edge servers Coexistence: }
To support a more diverse set of emerging services in the 5G network, the cloud datacenter and edge servers should coexist and be complementary to each other. However, some parts of a service may be executed at the mobile device itself, edge servers, or cloud datacenter. Given the available infrastructure and resource requirement of the service, identifying which part of the service to offload onto edge server/cloud datacenter and which not is a critical task. Further research is required to find intelligent strategies for coexistent cloud datacenter and edge server systems under realistic network conditions.

\textbf{Security and Privacy: }
In ECD, mobile devices upload the collected contents, which commonly contain sensitive and private information such as personal clinical data and business financial records, to the cache pool rather than the cloud datacenter. Therefore, such contents should be properly preprocessed on the cache pools before migrating between different cache pools or uploading to the cloud datacenter.

\section{Conclusion}
In this article, we have proposed an edge content delivery framework (ECD) based on mobile edge computing, which can effectively reduce the load on the core network and improve the QoE of users. A content server and a cache pool are introduced at the network edge, which is to store raw content generated by mobile users and store popular contents requested by mobile users, respectively. As a result, mobile users only need to upload all the collected contents to the content server rather than the cloud datacenter, which can significantly reduce the load on the core network. Furthermore, we have proposed an edge content delivery scheme, which prioritizes the top ranking cache pools for higher priority contents to reduce the response time; and an edge content update scheme, which update the contents in cache pools and cloud datacenter, based on the time-varying content polarity. The case study have demonstrated that the ECD can significant reduce the cost, compared with CDN. For the future work, we will study the location-dependent features when storing and updating contents.
\appendices

\ifCLASSOPTIONcaptionsoff
  \newpage
\fi

\end{document}